\documentstyle[buckow1,12pt]{article}
\begin{document}
\newcommand{\beq}{\begin{equation}}
\newcommand{\eeq}{\end{equation}}
\newcommand\sct[1]  {\mbox{$ $}\\[-.5em]{\bf #1.}\\[.16em]}

\rightline{HUB-EP-98-07}
\rightline{hep-th/9801156}

\input epsf
\newcount\figno
\figno=0
\def\fig#1#2#3{
\par\begingroup\parindent=0pt\leftskip=1cm\rightskip=1cm\parindent=0pt
\baselineskip=11pt
\global\advance\figno by 1
\epsfxsize=#3
\centerline{\epsfbox{#2}}
\vskip 12pt
{\bf Figure \the\figno:} #1\par
\endgroup\par
}
\def\figlabel#1{\xdef#1{\the\figno 
\mbox{ }}}
\def\encadremath#1{\vbox{\hrule\hbox{\vrule\kern8pt\vbox{\kern8pt
\hbox{$\displaystyle #1$}\kern8pt}
\kern8pt\vrule}\hrule}}

\def\titleline{Six-Dimensional Fixed Points from Branes }
\footnote{Based on
a talk given at the ``31st International Symposium Ahrenshoop
on the Theory of Elementary Particles''
Buckow, September 2-6, 1997.}

\def\authors{
Ilka Brunner, Andreas Karch
}

\def\addresses{
Institut f\"ur Physik, Humboldt-Universit\"at, \\
Invalidenstr. 110, D-10115 Berlin\\
}

\def\abstracttext{
We review the construction
of six dimensional $N=1$ fixed points in a brane picture
involving D6 branes stretching between NS 5 branes.
}

\makefront
\sct{Introduction: The classical Hanany-Witten brane setup}
Hanany and Witten \cite{hw} invented a brane configuration in IIB theory
to study 3 dimensional $N=4$ supersymmetric gauge theories.
The basic ingredients they used are IIB NS 5 branes, D5 branes
and D3 branes in flat space. The worldvolumes of the branes
are in the following directions 
\begin{center}
\vspace{.3cm}
\begin{tabular}{c|cccccccccc}
&$x^0$&$x^1$&$x^2$&$x^3$&$x^4$&$x^5$&$x^6$&$x^7$&$x^8$&$x^9$\\
\hline
NS 5&0&1&2&3&4&5&-&-&-&-\\
D 5&0&1&2&-&-&-&-&7&8&9\\
D 3&0&1&2&-&-&-&6&-&-&-\\
\end{tabular}
\vspace{.3cm}
\end{center}
It can be checked that this configuration preserves 1/4 of the
supersymmetries, that is 8 supercharges, as required for $N=4$
in $d=3$. The 3 branes are suspended between the 5 branes in the
$x_6$ direction. 
The 3d, $N=4$
theory is  realized in the directions which are common to all
branes. The point of view
we take is that the 5 branes are much heavier than the 3 branes
because they have two extra dimensions. The low energy dynamics is 
determined by the lowest dimensional brane in the setup. 
Moving
around the 3 brane corresponds to changing the moduli of
our theory. If we move around
5 branes this corresponds to changing parameters 
like masses, coupling constants, FI-terms.

What is the field theory on a 3 brane suspended between two
NS 5 branes? On an infinite 3 brane there is a theory of a
vector multiplet $V_{N=8}$ with 16
supercharges. The multiplet contains a vector and scalars corresponding
to the fluctuations of the 5 brane in the transversal directions.
In our setup, the $x_6$ direction is finite, therefore we are
left with a $2+1$ dimensional field theory. The effect of the
NS 5 branes is that SUSY  is broken to $N=4$.
The 3 brane motion in the $7,8,9$--direction is locked because the 3 brane
positions in these directions have to agree with the 5 brane positions.
Also, the scalar $A_6$ coming from the dimensional reduction of
the vector is projected out by the boundary conditions on the
NS 5 brane. 
Altogether, we are left with an $N=4$ vector multiplet.
To enhance the gauge group to $U(N_c)$ we can put $N_c$ 3 branes
on top of each other. The gauge coupling of the theory is 
related to the distance between the NS 5branes: 
$$
\frac{1}{g_{YM}^2} = \frac{\Delta x_6}{g_s}
$$
Here, $g_s$ is the string coupling and $1/g_s$ would be the gauge
coupling on an infinite D3.
We can include matter multiplets in the fundamental representation
by putting D5 branes in between the NS branes. Strings can then
stretch between the D5 and D3 branes yielding  matter in the fundamental
representation. The mass of these multiplets corresponds to the
distance (in the 345 direction) between the D5 and D3 branes.
An alternative way to include matter is to add semi-infinite 3 branes
to the left and right of the NS branes. Matter arises from strings
stretching between a semi-infinite and finite piece of the 3 brane.
The two descriptions of matter multiplets are related by the
Hanany Witten effect: We move the D5 branes off to infinity. When
they cross an NS brane a new D3 brane is created, which ends on
the NS brane.

The Hanany Witten setup can be T-dualized along the directions 3,4,5.
The dimensions of the D brane stretched between the two NS branes
increases in each step. This enables us to study theories with
8 supercharges in various dimensions.

\sct{Bending and RR charge conservation}
In higher dimensions it becomes very important to take into account
the disturbance caused by the D branes ending on the NS branes \cite{witten}.
The end of a Dd brane looks like a magnetic monopole in the worldvolume
of the NS 5 brane or as a charged particle on the $ 6-d$ dimensional
subspace transverse to the Dd brane. Dd branes ending from different
sides on the NS brane contribute with opposite charge. The consequence
is that the NS branes do not have a definite $x_6$ position, but
the $x_6$  coordinate obeys a Laplace equation:
$$
\Delta x_6(y) = 0,
$$
where $y$ parametrizes the transversal space. The ``true'' $x_6$
coordinate of the NS brane is  the $x_6$ value far away from
the disturbance. We can analyze the behaviour in various
dimensions by looking at the solutions to the Laplace equation
in various dimensions. In three dimensions, the case analyzed in
the previous paragraph, the solution behaves as
$$
x_6 = \frac{1}{|y|} \,+ \,constant,
$$
such that for $|y| \to \infty$ we get a definite value, which we
can call the $x_6$ position of the 5 brane.

In $d=4$ we obtain a logarithmic behaviour. The distance in $x_6$
between two NS branes is proportional to the 4 dimensional gauge
coupling, which is known to diverge logarithmically in 4 dimensions.
This is reproduced in the brane picture.

In 5 dimensions, the transversal space is one dimensional and we
obtain a linear bending of the NS brane. This can also be seen
from the fact that RR charge has to be conserved at the vertices
where different 5 branes come together. If we characterize a 5
brane by its charge under the NS and RR 2 forms, then a D brane
has charge (0,1) and an NS brane (1,0). If they end on each other,
a (1,1) brane emerges from the vertex.

\sct{The basic 6d brane setup}
We now want to study what happens in 6 dimensions in some more
detail \cite{d6, d62, zaff2}. 
We consider D6 branes stretching between NS branes. 
\vspace{0.2cm}
\fig{The brane configuration under consideration, giving rise
to a 6 dimensional field theory. Horizontal lines represent
D6 branes, the crosses represent NS 5 branes.}
{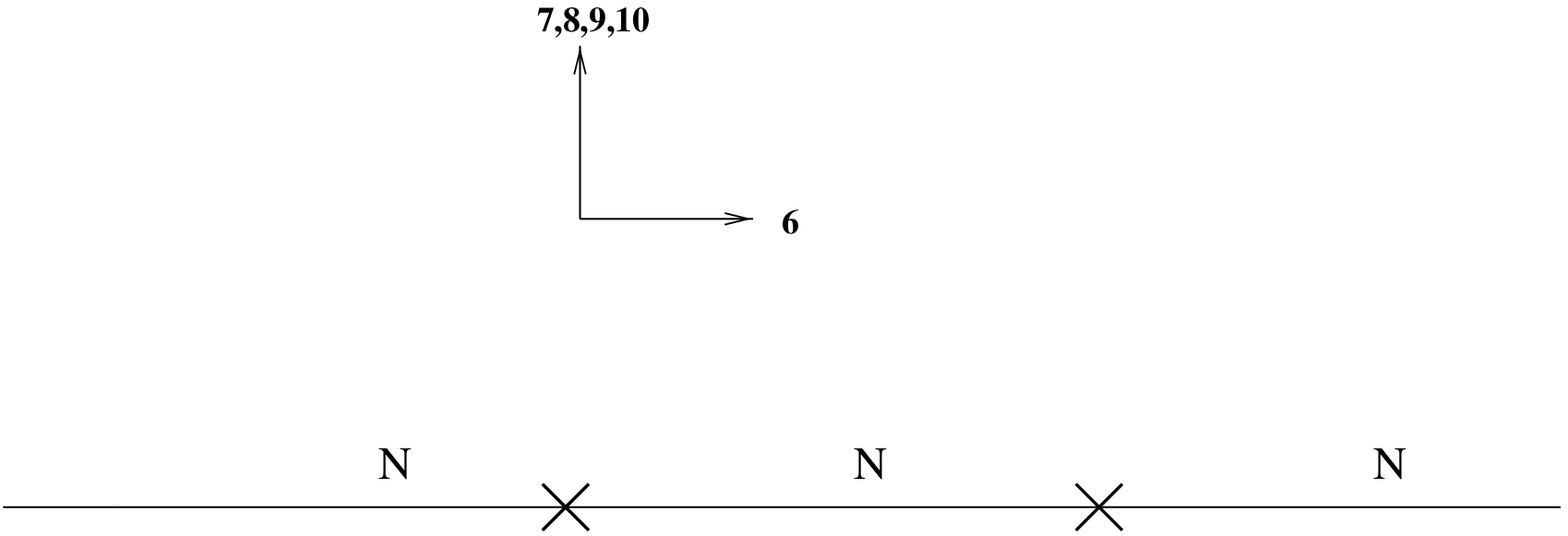}{10truecm}
\figlabel{\basic}
\vspace{0.2cm}
The configuration is shown in figure \basic.
The
worldvolume of the NS brane lies completely inside the worldvolume
of the D6 which ends on it. We include matter by semi-infinite
D6 branes extending to both sides of the NS branes and will discuss
D8 branes later on. Because there are no transversal directions of
the NS 5 brane left, there is no room for bending. The RR charge has
to cancel exactly at each vertex. The net charge is given by the
number of D6 branes ending from one side minus the number of D6 branes
ending from the other side. Thus, we only get a consistent picture
if:
$$
N_c = n_l = n_r,
$$
where $n_l$ ($n_r$) denotes the number of D6 ending from the left (right).
the total number of flavor giving semi infinite D6 is therefore
$$
N_f = n_l + n_r = 2 N_c
$$

\sct{The low energy field theory}
What is the low energy field theory interpretation of this brane setup?
According to the above philosophy, we have to look for the lowest dimensional
brane in the setup. In our setup, the NS branes are as light as the
finite D6 brane pieces. Only the semi infinite D6 branes are heavy.
This is different from brane configurations leading to lower dimensional
field theories, where the NS branes could always be considered as heavy
and their motions determined parameters in the theory.
The theory on a IIA NS 5 brane is the theory of a (0,2)-- tensor
multiplet. This multiplet consists of a tensor and 5 scalars (and
fermions). Because of the presence of the D6 branes, one half of
the SUSY is broken and we are left with a (0,1) theory. The tensor
multiplet decomposes into a (0,1) tensor, which only contains one
scalar, and a hypermultiplet, which contains 4 scalars. The 
hypermultiplet is projected out from the massless spectrum because
the position of the semi-infinite D6 branes fixes the position
of the NS branes, so that fluctuations in the transversal directions
are suppressed. The scalar in the tensor multiplet corresponds
to motions of the 5 branes in the $x_6$ direction. We have two
NS 5 branes and therefore two tensor multiplets, but effectively
we keep only one of them because one of the scalars can be taken to
describe the center of mass motion of the system. The vev of the
other scalar gives us the distance between the NS branes. It is 
therefore related to the coupling of the theory. If the two
5 branes come together we arrive at a strong coupling fixed points.
This theory contains tensionless strings coming from 
virtual membranes stretching between the 5 branes. 

Altogether, the branes describe an SU($N_c$) theory with a tensor
and $N_f$ hypers. The brane analysis gives the result that for
a consistent theory the number of fundamentals has to be
$ N_f = 2N_c$. It predicts a strong coupling fixed point with this
matter content. 

\sct{Inclusion of D8 branes}
So far, we included fundamental matter multiplets by semi-infinite
D6 branes. It should also be possible to describe the matter content
by higher dimensional D branes between the NS branes \cite{zaff}. In our case,
these are D8 branes. D8 branes are charged under a RR nine form
potential. The dual field strength of it is a constant. This constant
is related to th cosmological constant appearing in massive IIA 
supergravity. The D8 branes divide space-time into different regions
with different cosmological constant.
Whenever we cross a D8 brane, the cosmological constant jumps
by one unit. This is important for our brane configuration because
there is a term in the action of massive IIA supergravity which 
is proportional to the IIA mass parameter and which couples the NS
two form field $B$ to the field strength of the 7 form potential
under which the D6 is charged. The coupling reads:
$$
-m \int d^{10} x B \wedge * F^{(2)}
$$
($F^{(2)}$ is the field strength of the dual one form potential.)
This modifies the equations of motion for the 7 form potential
and therefore the RR charge cancellation condition. 
If a D6 brane ends on an NS brane, the equations of motion for
the 7 form potential, or equivalently the Bianchi identity for
the dual two form field strength is
$$
dF^{(2)} = d * F^{(8)} = \theta(x_7) \delta^{(456)} - mH,
$$
where $H=dB$. The $\delta$ term is the source term coming from the
D6 brane ending on the NS brane.
In the presence
of $m$ D8 branes the number of D6 branes ending from left and right
on the NS branes should differ by $m$. In this way, a D8 placed in
between the two NS branes has the same effect as a semi infinite
D6 ending on the NS brane.

\sct{ The field theoretical point of view}
We have seen that the brane construction leads via
RR charge conservation to  a restriction on
the number of vectors and hypers in the theory. This restriction
can be reproduced from a field theory point of view by an analysis
of the gauge anomaly. In six dimensions the anomaly arising
from vectors and hypers is
$$
I= {\rm tr}_{adj} F^4 - \sum_R n_R {\rm tr} F^4 
$$
$R$ denotes the representation of the matter multiplets. In our case
we only have $N_f$ fundamental matter multiplets. 
We can convert the trace in the adjoint
to a trace in the fundamental representation and obtain:
$$
I = (a-N_f) {\rm tr}_f F^4 + c ({\rm tr}_f F^2)^2
$$
where $a$ and $c$ are group dependent factors.
If the theory is consistent, the anomaly has to be cancelled. There are
various options: In the simplest case, both the prefactor of the $F^4$ term
and $c$ vanish. In this case the theory is certainly consistent. 
If only the prefactor of the $F^4$ term vanishes but $c$ does not vanish,
the anomaly can be cancelled by introducing a Green Schwarz tree-level
counterterm. This counterterm involves an antisymmetric tensor field.
In the case $c>0$ the anomaly can be cancelled without introducing
gravity, whereas in the case $c<0$ we can only get a consistent theory
if we introduce gravity. Note that in 6d a tensor can be divided
into a self dual and an anti self dual part. One piece is contained
in the gravity multiplet and the other in a tensor multiplet.
The tensor multiplet furthermore contains a scalar $\phi$, whereas
the gravity multiplet does not contain scalars.
In the remaining case that the prefactor of the $F^4$ term does not
vanish, the anomaly cannot be cancelled. Our gauge group is $SU(N_c)$.
Here, $a=2N_c$ and  $c=6$, so that we are in the situation, where
the anomaly can be cancelled by introducing a tensor, if $N_f = 2 N_c$.
This is precisely the result obtained from the brane picture.
For $N_c \leq 3$,  SU($N_c$) does not have an independent fourth order
Casimir. In this case, the condition $c>0$ only imposes an upper bound
on the number of flavors. Global anomalies restrict the possible matter
content further in these cases \cite{vafa}. 
The only additional possibilities for
SU(2) is 10 flavors, which can be realized by introducing an orientifold
and using SU(2) $\sim$ Sp(2). The additional possibilities in the SU(3)
case are 0 and 12 flavors and can not be seen in the brane picture.

From the brane picture we have predicted a strong coupling fixed point,
when the expectation value of the scalar in the tensor multiplet
vanishes. To verify this from a field theory point of view, we look
at the following part of the action:
$$
\frac{1}{g^2} {\rm tr} F_{\mu \nu}^{2} + \sqrt{c} \phi {\rm tr} F_{\mu \nu}^2
$$
We see, that one can absorb the bare gauge coupling into the expectation
value of the tensor to get an effective coupling
$$
\frac{1}{g_{eff}^2} = \sqrt{c} \phi
$$
This is the effective coupling we see in the brane picture.
At $\phi =0 $ there is a strong coupling fixed point \cite{seiberg}.

\sct{Modifications of the basic setup}
We can introduce further building blocks into our basic brane
setups to obtain other 6 dimensional fixed points.
Of course, we can build a chain of k NS branes connected by D6 branes,
leading to a product of SU gauge groups. This leads to a gauge 
group $SU(N_c)^k$ with bifundamentals and $2 N_c$ fundamentals.
We can also introduce orientifold 6 planes parallel to the D6 branes,
leading to Sp or SO gauge groups, depending on the sign of the
orientifold projection. The charge of the orientifold is twice
or minus twice the charge of a D6 brane. Furthermore, it changes
sign when it passes the NS branes. Taking this into account, RR
charge conservation leads for one group factor to the condition $N_f = N_c -8$
for SO and $N_f = N_c + 8$ for Sp. This is in agreement with
anomaly cancellation. 
Combining these two options we get gauge groups 
$ \{Sp(2N_c) \times SO(2N_c +8)\}^{k+1/2}$
Furthermore, we can introduce O8 planes parallel to flavor giving D8 branes.
If an NS 5 is stuck to the D8, we get SU gauge groups with an antisymmetric
or symmetric tensor (depending on the charge of the O8).

Many such possibilities have been studied in \cite{d62, zaff2}. The resulting
theories can also be obtained from 5 branes at ALE singularities
\cite{intblum}.

\end{document}